\documentclass[12pt,preprint]{aastex}

\usepackage{graphicx}
\usepackage{lineno}
\usepackage{amssymb,amsmath}

\slugcomment{ApJ Letters Accepted 17 March 2014}

\shorttitle{Deep Broadband Observations of the Distant Very High Energy Blazar PKS\,1424+240}
\shortauthors{Furniss et al.}

\begin{document}


\title{Deep Broadband Observations of the Distant Gamma-ray Blazar PKS\,1424+240}

\author{
S.~Archambault\altaffilmark{1},
T.~Aune\altaffilmark{2},
B.~Behera\altaffilmark{3},
M.~Beilicke\altaffilmark{4},
W.~Benbow\altaffilmark{5},
K.~Berger\altaffilmark{6},
R.~Bird\altaffilmark{7},
J.~Biteau\altaffilmark{8},
V.~Bugaev\altaffilmark{4},
K.~Byrum\altaffilmark{9},
J.~V~Cardenzana\altaffilmark{10},
M.~Cerruti\altaffilmark{5},
X.~Chen\altaffilmark{11,3},
L.~Ciupik\altaffilmark{12},
M.~P.~Connolly\altaffilmark{13},
W.~Cui\altaffilmark{14},
J.~Dumm\altaffilmark{15},
M.~Errando\altaffilmark{16},
A.~Falcone\altaffilmark{17},
S.~Federici\altaffilmark{3,11},
Q.~Feng\altaffilmark{14},
J.~P.~Finley\altaffilmark{14},
H.~Fleischhack\altaffilmark{3},
L.~Fortson\altaffilmark{15},
A.~Furniss\altaffilmark{28,*},
N.~Galante\altaffilmark{5},
G.~H.~Gillanders\altaffilmark{13},
S.~Griffin\altaffilmark{1},
S.~T.~Griffiths\altaffilmark{18},
J.~Grube\altaffilmark{12},
G.~Gyuk\altaffilmark{12},
D.~Hanna\altaffilmark{1},
J.~Holder\altaffilmark{6},
G.~Hughes\altaffilmark{3},
T.~B.~Humensky\altaffilmark{19},
C.~A.~Johnson\altaffilmark{8},
P.~Kaaret\altaffilmark{18},
M.~Kertzman\altaffilmark{20},
Y.~Khassen\altaffilmark{7},
D.~Kieda\altaffilmark{21},
H.~Krawczynski\altaffilmark{4},
F.~Krennrich\altaffilmark{10},
S.~Kumar\altaffilmark{6},
M.~J.~Lang\altaffilmark{13},
A.~S~Madhavan\altaffilmark{10},
G.~Maier\altaffilmark{3},
A.~McCann\altaffilmark{22},
K.~Meagher\altaffilmark{23},
P.~Moriarty\altaffilmark{24},
R.~Mukherjee\altaffilmark{16},
D.~Nieto\altaffilmark{19},
A.~O'Faol\'{a}in de Bhr\'{o}ithe\altaffilmark{7},
R.~A.~Ong\altaffilmark{2},
A.~N.~Otte\altaffilmark{23},
N.~Park\altaffilmark{25},
M.~Pohl\altaffilmark{11,3},
A.~Popkow\altaffilmark{2},
H.~Prokoph\altaffilmark{3},
J.~Quinn\altaffilmark{7},
K.~Ragan\altaffilmark{1},
J.~Rajotte\altaffilmark{1},
L.~C.~Reyes\altaffilmark{26},
P.~T.~Reynolds\altaffilmark{27},
G.~T.~Richards\altaffilmark{23},
E.~Roache\altaffilmark{5},
G.~H.~Sembroski\altaffilmark{14},
K.~Shahinyan\altaffilmark{15},
D.~Staszak\altaffilmark{1},
I.~Telezhinsky\altaffilmark{11,3},
J.~V.~Tucci\altaffilmark{14},
J.~Tyler\altaffilmark{1},
A.~Varlotta\altaffilmark{14},
V.~V.~Vassiliev\altaffilmark{2},
S.~Vincent\altaffilmark{3},
S.~P.~Wakely\altaffilmark{25},
A.~Weinstein\altaffilmark{10},
R.~Welsing\altaffilmark{3},
A.~Wilhelm\altaffilmark{11,3},
D.~A.~Williams\altaffilmark{8}
(The VERITAS Collaboration)\\
and\\
M.~Ackermann\altaffilmark{29}, 
M.~Ajello\altaffilmark{30}, 
A.~Albert\altaffilmark{31}, 
L.~Baldini\altaffilmark{32}, 
D.~Bastieri\altaffilmark{33,34}, 
R.~Bellazzini\altaffilmark{32}, 
E.~Bissaldi\altaffilmark{35}, 
J.~Bregeon\altaffilmark{36}, 
R.~Buehler\altaffilmark{29}, 
S.~Buson\altaffilmark{33,34}, 
G.~A.~Caliandro\altaffilmark{31,37}, 
R.~A.~Cameron\altaffilmark{31}, 
P.~A.~Caraveo\altaffilmark{38}, 
E.~Cavazzuti\altaffilmark{39}, 
E.~Charles\altaffilmark{31}, 
J.~Chiang\altaffilmark{31}, 
S.~Ciprini\altaffilmark{39,40}, 
R.~Claus\altaffilmark{31}, 
S.~Cutini\altaffilmark{39,40}, 
F.~D'Ammando\altaffilmark{41}, 
A.~de~Angelis\altaffilmark{42}, 
F.~de~Palma\altaffilmark{43,44}, 
C.~D.~Dermer\altaffilmark{45}, 
S.~W.~Digel\altaffilmark{31}, 
L.~Di~Venere\altaffilmark{43}, 
P.~S.~Drell\altaffilmark{31}, 
C.~Favuzzi\altaffilmark{43,44}, 
A.~Franckowiak\altaffilmark{31}, 
P.~Fusco\altaffilmark{43,44}, 
F.~Gargano\altaffilmark{44}, 
D.~Gasparrini\altaffilmark{39,40}, 
N.~Giglietto\altaffilmark{43,44}, 
F.~Giordano\altaffilmark{43,44}, 
M.~Giroletti\altaffilmark{41}, 
I.~A.~Grenier\altaffilmark{46}, 
S.~Guiriec\altaffilmark{47,48}, 
T.~Jogler\altaffilmark{31}, 
M.~Kuss\altaffilmark{32}, 
S.~Larsson\altaffilmark{49,50,51}, 
L.~Latronico\altaffilmark{52}, 
F.~Longo\altaffilmark{53,54}, 
F.~Loparco\altaffilmark{43,44}, 
P.~Lubrano\altaffilmark{55,56}, 
G.~M.~Madejski\altaffilmark{31}, 
M.~Mayer\altaffilmark{29}, 
M.~N.~Mazziotta\altaffilmark{44}, 
P.~F.~Michelson\altaffilmark{31}, 
T.~Mizuno\altaffilmark{57}, 
M.~E.~Monzani\altaffilmark{31}, 
A.~Morselli\altaffilmark{58}, 
S.~Murgia\altaffilmark{59}, 
E.~Nuss\altaffilmark{36}, 
T.~Ohsugi\altaffilmark{57}, 
J.~F.~Ormes\altaffilmark{60}, 
J.~S.~Perkins\altaffilmark{47}, 
F.~Piron\altaffilmark{36}, 
G.~Pivato\altaffilmark{34}, 
S.~Rain\`o\altaffilmark{43,44}, 
M.~Razzano\altaffilmark{32,61}, 
A.~Reimer\altaffilmark{62,31}, 
O.~Reimer\altaffilmark{62,31}, 
S.~Ritz\altaffilmark{63}, 
M.~Schaal\altaffilmark{64}, 
C.~Sgr\`o\altaffilmark{32}, 
E.~J.~Siskind\altaffilmark{65}, 
P.~Spinelli\altaffilmark{43,44}, 
H.~Takahashi\altaffilmark{66}, 
L.~Tibaldo\altaffilmark{31}, 
M.~Tinivella\altaffilmark{32}, 
E.~Troja\altaffilmark{47,67}, 
G.~Vianello\altaffilmark{31}, 
M.~Werner\altaffilmark{62}, 
M.~Wood\altaffilmark{31}
(The Fermi LAT Collaboration)
}

\altaffiltext{*}{Corresponding author}
\altaffiltext{1}{Physics Department, McGill University, Montreal, QC H3A 2T8, Canada}
\altaffiltext{2}{Department of Physics and Astronomy, University of California, Los Angeles, CA 90095, USA}
\altaffiltext{3}{DESY, Platanenallee 6, 15738 Zeuthen, Germany}
\altaffiltext{4}{Department of Physics, Washington University, St. Louis, MO 63130, USA}
\altaffiltext{5}{Fred Lawrence Whipple Observatory, Harvard-Smithsonian Center for Astrophysics, Amado, AZ 85645, USA}
\altaffiltext{6}{Department of Physics and Astronomy and the Bartol Research Institute, University of Delaware, Newark, DE 19716, USA}
\altaffiltext{7}{School of Physics, University College Dublin, Belfield, Dublin 4, Ireland}
\altaffiltext{8}{Santa Cruz Institute for Particle Physics and Department of Physics, University of California, Santa Cruz, CA 95064, USA}
\altaffiltext{9}{Argonne National Laboratory, 9700 S. Cass Avenue, Argonne, IL 60439, USA}
\altaffiltext{10}{Department of Physics and Astronomy, Iowa State University, Ames, IA 50011, USA}
\altaffiltext{11}{Institute of Physics and Astronomy, University of Potsdam, 14476 Potsdam-Golm, Germany}
\altaffiltext{12}{Astronomy Department, Adler Planetarium and Astronomy Museum, Chicago, IL 60605, USA}
\altaffiltext{13}{School of Physics, National University of Ireland Galway, University Road, Galway, Ireland}
\altaffiltext{14}{Department of Physics, Purdue University, West Lafayette, IN 47907, USA }
\altaffiltext{15}{School of Physics and Astronomy, University of Minnesota, Minneapolis, MN 55455, USA}
\altaffiltext{16}{Department of Physics and Astronomy, Barnard College, Columbia University, NY 10027, USA}
\altaffiltext{17}{Department of Astronomy and Astrophysics, 525 Davey Lab, Pennsylvania State University, University Park, PA 16802, USA}
\altaffiltext{18}{Department of Physics and Astronomy, University of Iowa, Van Allen Hall, Iowa City, IA 52242, USA}
\altaffiltext{19}{Physics Department, Columbia University, New York, NY 10027, USA}
\altaffiltext{20}{Department of Physics and Astronomy, DePauw University, Greencastle, IN 46135-0037, USA}
\altaffiltext{21}{Department of Physics and Astronomy, University of Utah, Salt Lake City, UT 84112, USA}
\altaffiltext{22}{Kavli Institute for Cosmological Physics, University of Chicago, Chicago, IL 60637, USA}
\altaffiltext{23}{School of Physics and Center for Relativistic Astrophysics, Georgia Institute of Technology, 837 State Street NW, Atlanta, GA 30332-0430}
\altaffiltext{24}{Department of Life and Physical Sciences, Galway-Mayo Institute of Technology, Dublin Road, Galway, Ireland}
\altaffiltext{25}{Enrico Fermi Institute, University of Chicago, Chicago, IL 60637, USA}
\altaffiltext{26}{Physics Department, California Polytechnic State University, San Luis Obispo, CA 94307, USA}
\altaffiltext{27}{Department of Applied Physics and Instrumentation, Cork Institute of Technology, Bishopstown, Cork, Ireland}
\altaffiltext{28}{Kavli Institute for Particle Astrophysics and Cosmology, SLAC National Accelerator Laboratory, Stanford University, Stanford, CA 94305, USA}

\altaffiltext{29}{Deutsches Elektronen Synchrotron DESY, D-15738 Zeuthen, Germany}
\altaffiltext{30}{Space Sciences Laboratory, 7 Gauss Way, University of California, Berkeley, CA 94720-7450, USA}
\altaffiltext{31}{W. W. Hansen Experimental Physics Laboratory, Kavli Institute for Particle Astrophysics and Cosmology, Department of Physics and SLAC National Accelerator Laboratory, Stanford University, Stanford, CA 94305, USA}
\altaffiltext{32}{Istituto Nazionale di Fisica Nucleare, Sezione di Pisa, I-56127 Pisa, Italy}
\altaffiltext{33}{Istituto Nazionale di Fisica Nucleare, Sezione di Padova, I-35131 Padova, Italy}
\altaffiltext{34}{Dipartimento di Fisica e Astronomia ``G. Galilei'', Universit\`a di Padova, I-35131 Padova, Italy}
\altaffiltext{35}{Istituto Nazionale di Fisica Nucleare, Sezione di Trieste, and Universit\`a di Trieste, I-34127 Trieste, Italy}
\altaffiltext{36}{Laboratoire Univers et Particules de Montpellier, Universit\'e Montpellier 2, CNRS/IN2P3, Montpellier, France}
\altaffiltext{37}{Consorzio Interuniversitario per la Fisica Spaziale (CIFS), I-10133 Torino, Italy}
\altaffiltext{38}{INAF-Istituto di Astrofisica Spaziale e Fisica Cosmica, I-20133 Milano, Italy}
\altaffiltext{39}{Agenzia Spaziale Italiana (ASI) Science Data Center, I-00133 Roma, Italy}
\altaffiltext{40}{Istituto Nazionale di Astrofisica - Osservatorio Astronomico di Roma, I-00040 Monte Porzio Catone (Roma), Italy}
\altaffiltext{41}{INAF Istituto di Radioastronomia, 40129 Bologna, Italy}
\altaffiltext{42}{Dipartimento di Fisica, Universit\`a di Udine and Istituto Nazionale di Fisica Nucleare, Sezione di Trieste, Gruppo Collegato di Udine, I-33100 Udine, Italy}
\altaffiltext{43}{Dipartimento di Fisica ``M. Merlin" dell'Universit\`a e del Politecnico di Bari, I-70126 Bari, Italy}
\altaffiltext{44}{Istituto Nazionale di Fisica Nucleare, Sezione di Bari, 70126 Bari, Italy}
\altaffiltext{45}{Space Science Division, Naval Research Laboratory, Washington, DC 20375-5352, USA}
\altaffiltext{46}{Laboratoire AIM, CEA-IRFU/CNRS/Universit\'e Paris Diderot, Service d'Astrophysique, CEA Saclay, 91191 Gif sur Yvette, France}
\altaffiltext{47}{NASA Goddard Space Flight Center, Greenbelt, MD 20771, USA}
\altaffiltext{48}{NASA Postdoctoral Program Fellow, USA}
\altaffiltext{49}{Department of Physics, Stockholm University, AlbaNova, SE-106 91 Stockholm, Sweden}
\altaffiltext{50}{The Oskar Klein Centre for Cosmoparticle Physics, AlbaNova, SE-106 91 Stockholm, Sweden}
\altaffiltext{51}{Department of Astronomy, Stockholm University, SE-106 91 Stockholm, Sweden}
\altaffiltext{52}{Istituto Nazionale di Fisica Nucleare, Sezione di Torino, I-10125 Torino, Italy}
\altaffiltext{53}{Istituto Nazionale di Fisica Nucleare, Sezione di Trieste, I-34127 Trieste, Italy}
\altaffiltext{54}{Dipartimento di Fisica, Universit\`a di Trieste, I-34127 Trieste, Italy}
\altaffiltext{55}{Istituto Nazionale di Fisica Nucleare, Sezione di Perugia, I-06123 Perugia, Italy}
\altaffiltext{56}{Dipartimento di Fisica, Universit\`a degli Studi di Perugia, I-06123 Perugia, Italy}
\altaffiltext{57}{Hiroshima Astrophysical Science Center, Hiroshima University, Higashi-Hiroshima, Hiroshima 739-8526, Japan}
\altaffiltext{58}{Istituto Nazionale di Fisica Nucleare, Sezione di Roma ``Tor Vergata", I-00133 Roma, Italy}
\altaffiltext{59}{Center for Cosmology, Physics and Astronomy Department, University of California, Irvine, CA 92697-2575, USA}
\altaffiltext{60}{Department of Physics and Astronomy, University of Denver, Denver, CO 80208, USA}
\altaffiltext{61}{Funded by contract FIRB-2012-RBFR12PM1F from the Italian Ministry of Education, University and Research (MIUR)}
\altaffiltext{62}{Institut f\"ur Astro- und Teilchenphysik and Institut f\"ur Theoretische Physik, Leopold-Franzens-Universit\"at Innsbruck, A-6020 Innsbruck, Austria}
\altaffiltext{63}{Santa Cruz Institute for Particle Physics, Department of Physics and Department of Astronomy and Astrophysics, University of California at Santa Cruz, Santa Cruz, CA 95064, USA}
\altaffiltext{64}{National Research Council Research Associate, National Academy of Sciences, Washington, DC 20001, resident at Naval Research Laboratory, Washington, DC 20375, USA}
\altaffiltext{65}{NYCB Real-Time Computing Inc., Lattingtown, NY 11560-1025, USA}
\altaffiltext{66}{Department of Physical Sciences, Hiroshima University, Higashi-Hiroshima, Hiroshima 739-8526, Japan}
\altaffiltext{67}{Department of Physics and Department of Astronomy, University of Maryland, College Park, MD 20742, USA}

\email{amy.furniss@gmail.com}

\begin{abstract}
We present deep VERITAS observations of the blazar PKS 1424+240, along with contemporaneous \textit{Fermi} Large Area Telescope, \textit{Swift} X-ray Telescope and \textit{Swift} UV Optical Telescope data between 2009 February 19 and 2013 June 8.  This blazar resides at a redshift of $z\ge0.6035$, displaying a significantly attenuated gamma-ray flux above 100~GeV due to photon absorption via pair-production with the extragalactic background light.  We present more than 100 hours of VERITAS observations from three years, a multiwavelength light curve and the contemporaneous spectral energy distributions.  The source shows a higher flux of (2.1$\pm0.3$)$\times10^{-7}$ ph m$^{-2}$s$^{-1}$ above 120 GeV in 2009 and 2011 as compared to the flux measured in 2013, corresponding to (1.02$\pm0.08$)$\times10^{-7}$ ph m$^{-2}$s$^{-1}$ above 120 GeV.  The measured differential very high energy (VHE; $E\ge100$ GeV) spectral indices are $\Gamma=$3.8$\pm$0.3, 4.3$\pm$0.6 and 4.5$\pm$0.2 in 2009, 2011 and 2013, respectively. No significant spectral change across the observation epochs is detected.  
We find no evidence for variability at gamma-ray opacities of greater than $\tau=2$, where it is postulated that any variability would be small and occur on longer than year timescales if hadronic cosmic-ray interactions with extragalactic photon fields provide a secondary VHE photon flux. The data cannot rule out such variability due to low statistics.  \end{abstract}


\section{Introduction}
PKS\,1424+240 (VER\,J1427+237) is a distant very high energy (VHE; $E\ge100$ GeV) blazar at $z\ge0.6035$ \citep{furniss1424}.  At this \textit{minimum} distance, the intrinsic VHE emission is expected to be significantly absorbed by the extragalactic background light (EBL) via pair-production, $\gamma + \gamma \rightarrow e^{+} + e^{-}$ \citep{nikishov}.  The absorption of VHE gamma rays by the EBL can be estimated using the model-dependent gamma-ray opacity, $\tau(E,z)$.  The source flux, $F_{\rm int}$, can be estimated from the observed flux, $F_{\rm obs}$, using the relation $F_{\rm int} = F_{\rm obs}\times e^{\tau(E,z)}$.  

The EBL cannot be directly measured due to foreground sources. The modification of distant VHE blazar spectra has been used to estimate the spectral properties of the EBL \citep{aharonian2006,albert2008}, providing photon density upper limits consistent with the observational lower limits set by galaxy counts \citep{werner}.  Recent work has indicated that the EBL density is closer to the lower limits than the upper limits \citep{HESSEBL,hornsmeyer,fermiEBL}.  The distance to PKS\,1424+240 makes the source ideal for studying extragalactic VHE photon propagation.  

The high-energy spectral energy distribution (SED) measured in initial observations by VERITAS and the \textit{Fermi} Large Area Telescope (LAT) \citep{acciari1424} is investigated in \cite{furniss1424}, showing an absorption-corrected spectrum suggestive of VHE spectral hardening, though not beyond the conservative $\Gamma=1.5$ spectral limitation (where $dN/dE \propto E^{-\Gamma}$) described in, e.g., \cite{aharonian2006}.  

In an effort to understand the gamma-ray emission from PKS\,1424+240, we analyze deeper observations by VERITAS and LAT, including more than four times the exposure in \cite{acciari1424} and \cite{furniss1424}.  In order to minimize hardening introduced from EBL absorption corrections, we explore the gamma-ray observations using the low-density ``fixed" model from \cite{gilmore2012}.  This model, also providing compatible fits to LAT data in \cite{fermiEBL}, is comparable with that of \cite{franceschini} used in \cite{HESSEBL}, and provides similar absorption-corrections as compared to other EBL models, e.g. \cite{kneiske,dominguez,finke}.  Luminosities calculated in this work use a H$_0$ = 100 $h$ km s$^{-1}$Mpc$^{-1}$ where $h $= 0.7.

\section{Observations and Results}
\subsection{VERITAS}
VERITAS comprises four imaging atmospheric Cherenkov telescopes and is sensitive to gamma rays between $\sim$100 GeV and $\sim$30 TeV \citep{holder2006}. The VERITAS observations of PKS\,1424+240 were performed during three years.  The first season (MJD 54881-55003) provides 28 hours of quality-selected livetime and is reanalyzed here, showing results consistent with those reported in \cite{acciari1424}.  The second season encompasses 14 quality-selected hours of observation between MJD 55598 and 55711, while the third season includes data spanning MJD 56334 to 56447, and provide 67 hours of quality-selected livetime with a low threshold of 100 GeV, enabled by a camera upgrade in 2012.   

The observations were taken at 0.5$^{\circ}$ offset in each of the four cardinal directions to enable simultaneous background estimation using the reflected-region method \citep{fomin}.  The recorded shower images are parameterized by their principal moments.   Selection criteria are applied to the values of mean scaled width (MSW), and mean scaled length (MSL), apparent altitude of the maximum Cherenkov emission (shower maximum), and $\theta$, the angular distance between the position of PKS\,1424+240 and the reconstructed origin of the event, giving an efficient suppression of the far more abundant cosmic-ray background.    The cuts applied to all data are MSW$<$1.1, MSL$<$1.3, shower maximum $>$7 km, and $\theta<0.17^{\circ}$.  These cuts were optimized \textit{a priori} to yield the highest sensitivity for a soft ($\Gamma\sim3.5$) source with 5\% of the Crab Nebula gamma-ray flux.\footnote{Flux calculated according to \cite{albert2008}.}  These cuts are different from those in \cite{acciari1424} because of improvements in the analysis software and detector simulation. The results are independently reproduced with two analysis packages \citep{cogan,heike} and are summarized in Table~1.  The same analysis was applied to data from the Crab Nebula for each season, providing compatible flux and spectral results with no evidence of an energy bias shift after the camera upgrade.   In particular, the integrated fluxes measured above 200 GeV agree to 11\% or better (1 $\sigma$ confidence).   The systematic uncertainty on the flux for a soft source like PKS\,1424+240 is estimated at $\sim40\%$ and is regarded as constant for each of the observing periods.

The 2009 and 2011 observations show the source to have a flux of 4.6\% of the Crab flux above 120 GeV, with indices of $\Gamma$=3.8$\pm$0.3 and 4.3$\pm$0.3, respectively.    The longer exposures obtained in 2009 and 2013 allow for the the reconstruction of a significant spectral point in a higher energy bin than is possible with the 2011 data (see Figure 1).  In an attempt to minimize a bias in the final spectral bin width, the energy binning is systematically determined, starting at 100 GeV with bins of equal logarithmic width, initially corresponding to 15 GeV.  The first bin that does not provide sufficient statistics for a spectral point ($<2$ standard deviations; $\sigma$), is doubled in width compared to the preceding bin size.  This wider bin is then utilized in the analysis to derive higher energy spectral points.  The first instance where the doubling procedure does not provide a significant detection is reported with a 99\% confidence level upper limit \citep{rolke}, assuming the same spectral index that fit to the preceding bins.  The spectral points are given at the energy corresponding to the event-weighted average in the bin. For the last bin, with bin edges 375 GeV and 750 GeV, the weighted average corresponds to 510 GeV.

During the 2013 observations the source was in a dimmer VHE state of 2.2\% Crab above 120 GeV (see Figure~1).  The VHE spectral index does not appear to change during this low state, displaying an index of $\Gamma=4.5\pm0.2$.  The observations over each season are shown in the top panel of the light curve (Figure 2).    The 2009 and 2013 observations show different states, with integral flux values above 120~GeV of (2.1$\pm$0.3)$\times10^{-7}$ph m$^{-2}$ s$^{-1}$ and  (1.02$\pm$0.08)$\times10^{-7}$ph m$^{-2}$ s$^{-1}$, respectively.  Additionally, a constant fit to the VHE light curve shows less than $1.1\times10^{-5}$ probability of a steady flux ($\chi^2=22.7$ with 2 degrees of freedom; DOF).  A search for variability above an opacity of $\tau=2$ (corresponding to 310 GeV according to the \citealt{gilmore2012} EBL model) does not provide significant evidence of variability given the very limited statistics at high energies, with integral flux values above 310 GeV of $(5.6\pm3.8)\times10^{-9}$m$^{-2}$ s$^{-1}$ for 2009/2011 combined data and $(3.6\pm1.8)\times10^{-9}$m$^{-2}$ s$^{-1}$ for 2013 data.

The power-law fit to the 2013 data is shown in Figure 3 with an envelope representing a $\pm$40\% systematic error on the flux convolved with a systematic error on the index of $\pm$0.3.   
The data are corrected for absorption by the EBL assuming the model from \cite{gilmore2012} at the minimum redshift of $z=0.6035$, resulting in a power-law fit ($\chi^2$/DOF=9.1/9, probability of 0.428) with index $\Gamma$=3.0$\pm$0.4. 
The 2009 absorption-corrected data provide a power-law fit with $\Gamma$=2.8$\pm$0.7 ($\chi^2$/DOF=4.7/6, probability of 0.583).  
As a consistency check, the data are also shown in Figure 3 with constant binning above 250 GeV.   None of the individual points above 400 GeV are statistically significant in this representation. 

\subsection{\textit{Fermi} LAT}
The \textit{Fermi} LAT is a pair-conversion telescope sensitive to photons between 20 MeV and several hundred GeV \citep{atwood}.  PKS\,1424+240 is a bright gamma-ray source first reported in \cite{abdo09b}. Multiple epochs of LAT data are analyzed, including the complete \textit{Fermi} LAT dataset up to the time of analysis (MJD 54682 to 56452) and time intervals selected to be contemporaneous with the VERITAS observations, summarized in Table~1.  The spectral parameters for the contemporaneous data are calculated using the unbinned maximum-likelihood method implemented in the LAT \texttt{ScienceTools} software package version \texttt{v9r31p1}, available from the Fermi Science Support Center.  The spectral parameters for the full dataset are calculated using the binned maximum-likelihood method.   Only events from the ``source" class with energy above 100 MeV within a 12$^{\circ}$ radius of PKS\,1424+240 with a zenith angle of $< 100^{\circ}$ and detected when the spacecraft rocking angle was $< 52^{\circ}$ are used.  All sources within 12$^{\circ}$ of the central source in the second LAT catalog  (2FGL, \citealt{2fgl}) are included in the model.  The normalizations of the components were allowed to vary freely during the spectral point fitting, which was performed using the instrument response function \texttt{P7SOURCE\_V6}.  The Galactic diffuse emission and an isotropic component, which is the sum of the extragalactic diffuse gamma-ray emission and the residual charged particle background, are modeled using the recommended files.\footnote{The files used were \texttt{gal\_2yearp7v6\_v0.fits} for the Galactic diffuse and \texttt{iso\_p7v6source.txt} for the isotropic diffuse component available at \tt http://fermi.gsfc.nasa.gov/ssc/data/access/lat/BackgroundModels.html.}  The flux systematic uncertainty is estimated as approximately $5\%$ at $560$\,MeV and under $10\%$ at $10$\,GeV and above.

The data are fit with power-law models for each of three contemporaneous epochs in 2009, 2011 and 2013, showing no significant variations (see Table~1).  The three epochs were also combined (referred to as ``Contemp." in Table~1) and fit with a power law.  Additionally, an extended LAT data set (MJD 54682 to 56452) is analyzed using the binned-likelihood method.  The larger dataset is fit with a curved log-parabolic model including EBL absorption with the \cite{gilmore2012} EBL model, since there is a significant (TS$>9$; \citealt{mattox}) detection up to 300 GeV.  The contemporaneous 2009 and 2013 data are shown in Figure 4.

The data above 1\,GeV are also analyzed in 28-day bins (see Figure~2).  This light curve displays variability, with a probability of $\sim1\times10^{-11}$ of being steady ($\chi^2$=159.7 with 57 DOF).  However, a search for variability above 10 GeV using the Bayesian Block method from \cite{scargle} with 1\% specified as the acceptable fraction of false positives shows no evidence of variability, in agreement with the lack of significant variability found above 10\,GeV in the \textit{Fermi} LAT hard sources catalog (1FHL; \citealt{fhl}).

\subsection{\textit{Swift} XRT}
The X-ray Telescope (XRT) onboard the \textit{Swift} satellite \citep{gehrels} is a focusing X-ray telescope sensitive to photons with energy between 0.2 and 10 keV.  Thirty observations of PKS\,1424+240 summing to 51 ks have been collected between 2009 June 11 and 2013 May 10 (MJD 54993 and 56422), inclusive.   Observations were taken in photon counting mode with the count rate ranging from 0.1 to 1.1 counts per second.  Pile-up effects are accounted for when the count rate exceeds 0.5 counts per second using an annular source region, with a 1-2 pixel inner radius and a 20 pixel outer radius.  The data are analyzed as described in \cite{burrows05} with HEASOFT\,6.9 and XSPEC version 12.6.0.

For spectral fitting, the photons are grouped by energy to require a minimum of 20 counts per bin, and fit with an absorbed power law (\textit{tbabs(po)}) between 0.3 and 10 keV, fixing the neutral hydrogen column density to $3\times10^{20}$ cm$^{-2}$, as quoted in \cite{kalberla}.  The data are also fit with an absorbed log-parabolic model (\textit{tbabs(logpar)}), finding curvature parameters consistent with zero.  Due to the lack of curvature, we only discuss the power-law fitted parameters here.  

The 2$-$10 keV integral flux values are derived for each observation and shown in Figure 2.  The X-ray light curve is clearly variable, with a constant fit giving a $\chi^2$ of 2200 for 30 DOF.  X-ray energy spectra are extracted for the highest and lowest states (from MJD 54997 and 56368, respectively).   The high and low flux states differ by a factor of $\sim10$ and have photon indices of $\alpha=2.36\pm0.04$ and $\alpha=2.8\pm0.1$, respectively.  These X-ray states correspond to 2$-$10 keV rest frame luminosities of at least $2.5 \times 10^{46}$ erg s$^{-1}$ and $2.4 \times 10^{45}$ erg s$^{-1}$, respectively, assuming $z=0.6035$.  In order to represent the intrinsically emitted SED, the spectra corrected for the column density absorption are shown in Figure 4.

\subsection{\textit{Swift} UVOT}
\textit{Swift}-UVOT exposures were taken with UVW1, UVM2, and UVW2 filters \citep{poole}.  The UVOT photometry is performed using \texttt{HEASoft} \textit{uvotsource}.  The circular source region has a $5\arcsec$ radius and the background region consists of several 15$\arcsec$ radii circles of nearby empty sky. The results are reddening-corrected using the E(B-V) coefficients in \cite{schlegel}.  The Galactic extinction coefficients are applied according to \cite{fitzpatrick}.  The uncertainty in the reddening E(B-V) is the largest source of error.  The UV light curve is shown in Figure~2, with the UV flux values corresponding to the high and low X-ray states plotted in Figure~4.  UV variability is apparent, with a pattern similar to the X-ray band. 

\section{Absorption-corrected Broadband SED}
Two broadband SEDs of PKS\,1424+240 are shown in Figure 4, corresponding to relatively high and low states.  Two inset plots show the absorption-corrected VHE data  according to the EBL model in \cite{gilmore2012}.  There is an indication of spectral hardening at the highest energies in the absorption-corrected VHE spectrum.  Similar 
results are seen when absorption correction is done according to a variety of EBL models, such as \cite{dominguez,finke,franceschini}.  

The contemporaneous LAT data are also shown in the insets of Figure 4, but the correction for EBL absorption is $<1\%$ at the highest energy LAT spectral points.  Spectral results derived from the full LAT observations are also shown, and are consistent with the VERITAS observations.  The synchrotron peaks are shown with \textit{Swift} XRT and UVOT observations from relatively high and low states.  Since the synchrotron peak is known to be above UV energies (e.g. \citealt{acciari1424}), these observations constrain the location of the synchrotron peak (10$^{15}$-10$^{16}$ Hz) during relatively low and high synchrotron flux states.  The source is not detected between 14 and 195 keV by the Burst Alert Telescope onboard \textit{Swift} in 70 months of data \citep{BAT}.

\section{Discussion}
The blazar PKS\,1424+240 resides at $z\ge0.6035$, with a VHE flux that is significantly attenuated by the EBL.  Discovery observations of this source by VERITAS have shown a marginal indication of spectral hardening at the highest energies, after correction by the EBL \citep{furniss1424}.   While a similar effect is seen in the deep observations obtained in 2013, the significance of the effect remains marginal because of the lower overall flux level during this epoch.  In both epochs the data are consistent with a simple power law, even after correction for absorption by the EBL.
If the indication of spectral hardening could be confirmed, one possible explanation would be an over-estimation of the EBL density, although the results shown use one of the lowest density EBL models currently available, which approaches the galaxy count lower limits of the EBL density at $z\sim0$.  

The possible spectral hardening is of great interest, because if it is not from over-estimation of the EBL, there are a number of physical mechanisms which can produce hardening with increasing energy. Second-order synchrotron self Compton emission, pair-cascades initiated by pion decay in hadronic emission scenarios \citep{boettcher} or internal photon-photon absorption \citep{aharonian2008} can produce hard components at high energy.  There are also scenarios that describe spectral hardening as arising from hadronic cosmic-ray line-of-sight interactions with the cosmic microwave background and EBL.  These processes can produce secondary gamma rays close to the observer, hardening the observed VHE spectrum \citep{essey2010a,essey2010b,essey2011,essey2012,murase2012,razzaque2012,prosekin2012,aharonian2013,zheng2013,kalashev2013,inoue2013}.  This component is expected to become dominant at high energies where the EBL opacity is greater than $\sim$2 and is not expected to vary on timescales shorter than about a year.  The VERITAS observations above 310 GeV (where $\tau=2$ according to \citealt{gilmore2012} and \citealt{kneiske}) do not show significant variability between 2009 and 2013, nor can they strongly exclude it.  More exotic theories, involving Lorentz invariance violation \citep{LIV} or axion-like particles (ALPs), might also produce spectral hardening at high energies, e.g. \cite{sanchezconde}.  

The blazar can be categorized as an high-synchrotron-peaked (HSP) BL Lac, with a synchrotron peak above 10$^{15}$ Hz \citep{abdoSED} and an isotropic luminosity above 400 GeV of 1.03$\times10^{44}$ erg s$^{-1}$.    At $z\ge0.6035$, it is apparent that PKS\,1424+240 represents a powerful tool for studying intrinsic emission mechanism(s) within blazar jets, extragalactic cosmic-ray propagation and the propagation of VHE photons across extragalactic space.  Future studies will benefit from additional VHE observations as well as from any additional information that will be obtained about the redshift, e.g. from HST/STIS UV observations.

\acknowledgments
This research is supported by grants from the U.S. Department of Energy Office of Science, the U.S. National Science Foundation and the Smithsonian Institution, by NSERC in Canada, by Science Foundation Ireland (SFI 10/RFP/AST2748) and by STFC in the U.K. We acknowledge the excellent work of the technical support staff at the Fred Lawrence Whipple Observatory and at the collaborating institutions in the construction and operation of the instrument.

Support for program HST-GO-12863 was provided by NASA, awarded through the Space Telescope Science Institute, operated by the Association of Universities for Research in Astronomy,  Inc., for NASA, under contract NAS 5-26555. 

The $Fermi$ LAT Collaboration acknowledges support from a number of agencies and institutes for both development and the operation of the LAT as well as scientific data analysis. These include NASA and DOE in the United States, CEA/Irfu and IN2P3/CNRS in France, ASI and INFN in Italy, MEXT, KEK, and JAXA in Japan, and the K.~A.~Wallenberg Foundation, the Swedish Research Council and the National Space Board in Sweden. Additional support from INAF in Italy and CNES in France for science analysis during the operations phase is also gratefully acknowledged.

{\it Facilities:} \facility{VERITAS}, \facility{\textit{Fermi}}, \facility{\textit{Swift}}.

\begin{figure}
\begin{center}
\epsscale{1.1}
\includegraphics[scale=0.85]{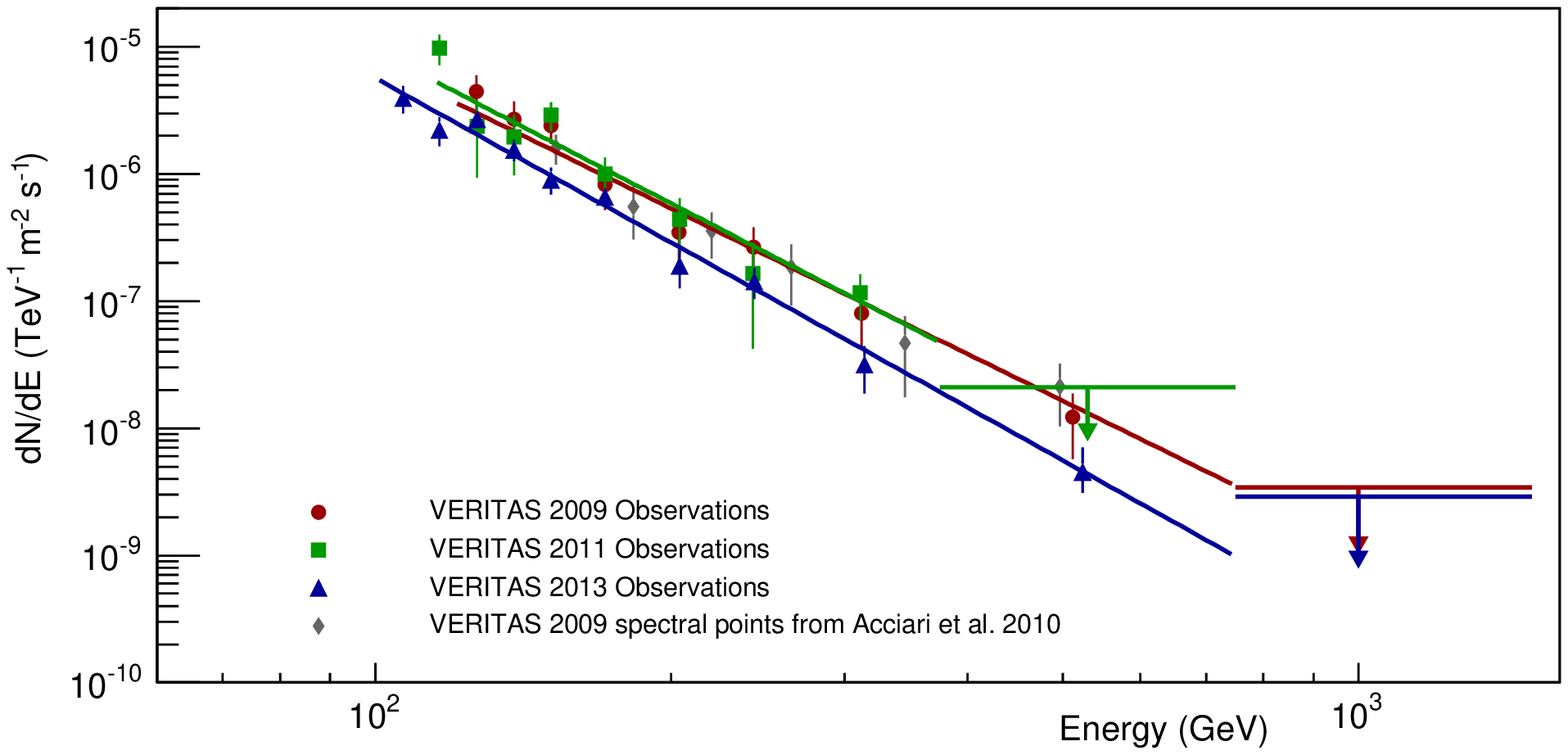}
\caption{Observed VHE spectra of PKS\,1424+240 derived from three years of VERITAS observation.  The 2009, 2011 and 2013 spectral results are shown with 1$\sigma$ error bars. The spectrum from \cite{acciari1424} is also shown in gray. See Table~1 for details.\label{fig1}}
\end{center}
\end{figure}

\begin{figure}
\begin{center}
\epsscale{1.1}
\includegraphics[scale=0.85]{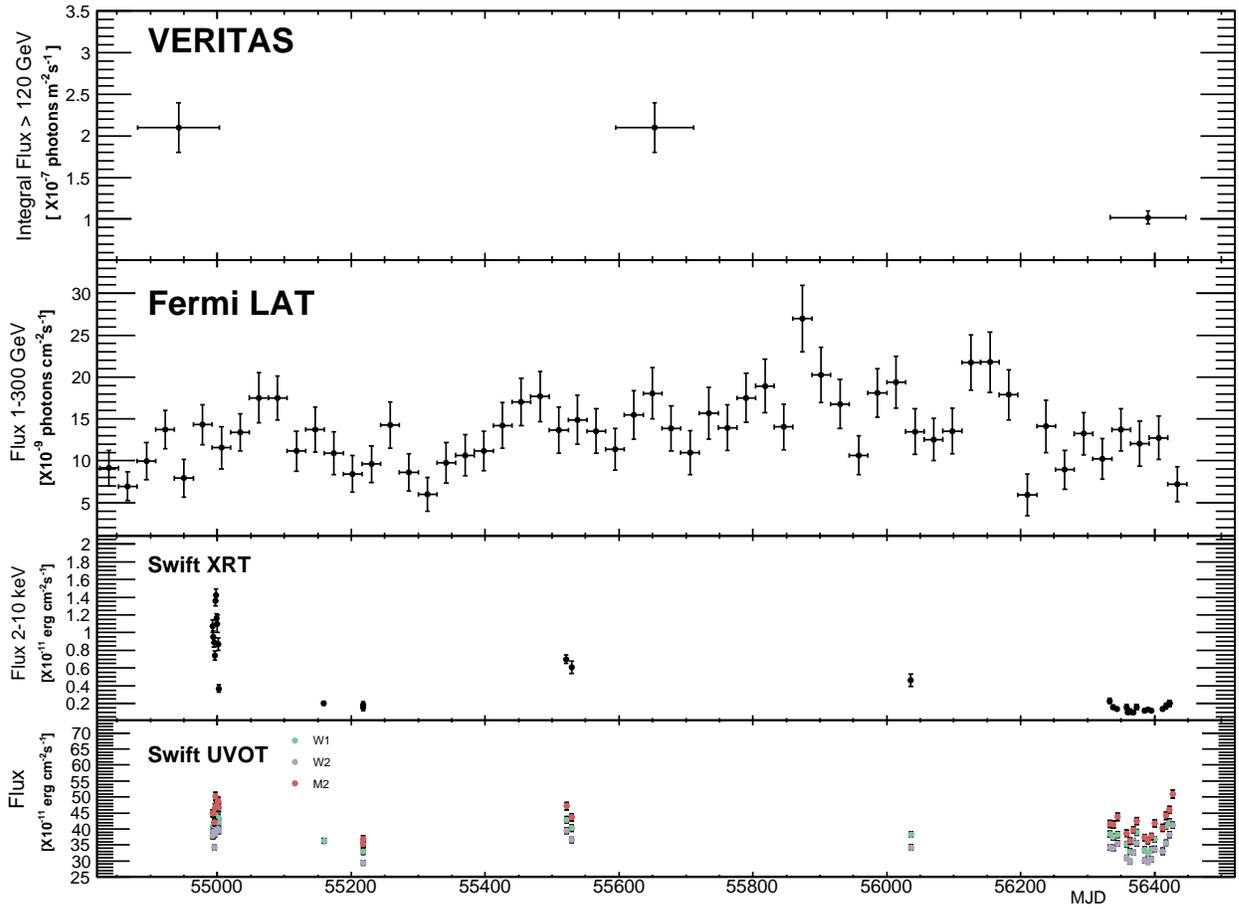}
\caption{VERITAS, \textit{Fermi} LAT and \textit{Swift} X-ray and UV light curves for PKS\,1424+240.\label{fig2}}
\end{center}
\end{figure}
 
 \begin{figure}
\begin{center}
\epsscale{1.1}
\includegraphics[scale=0.85]{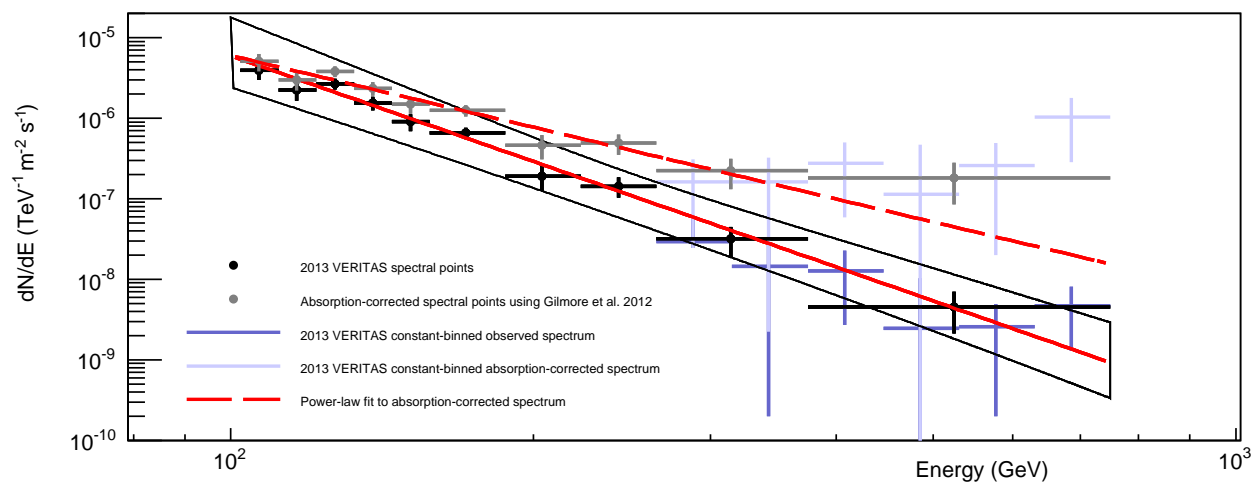}
\caption{The VHE spectrum derived from the 2013 VERITAS dataset, with a significant detection between 100 and 750 GeV (black points).  The solid red line represents the power-law fit to the observed data.  These data are shown with an envelope representing a 40\% systematic error on the source flux and index error of $\pm$0.3.  The data are also shown after correction for EBL-absorption by \cite{gilmore2012},  assuming $z=0.6035$ (gray points), along with the power-law (long-dashed line) fit to the absorption-corrected data.  In blue, spectral points above 250 GeV derived with constant binning are shown. None of the individual points above 400 GeV is statistically significant in this representation. \label{fig3}}
\end{center}
\end{figure}

\begin{figure}
\begin{center}
\epsscale{1.1}
\includegraphics[scale=0.75]{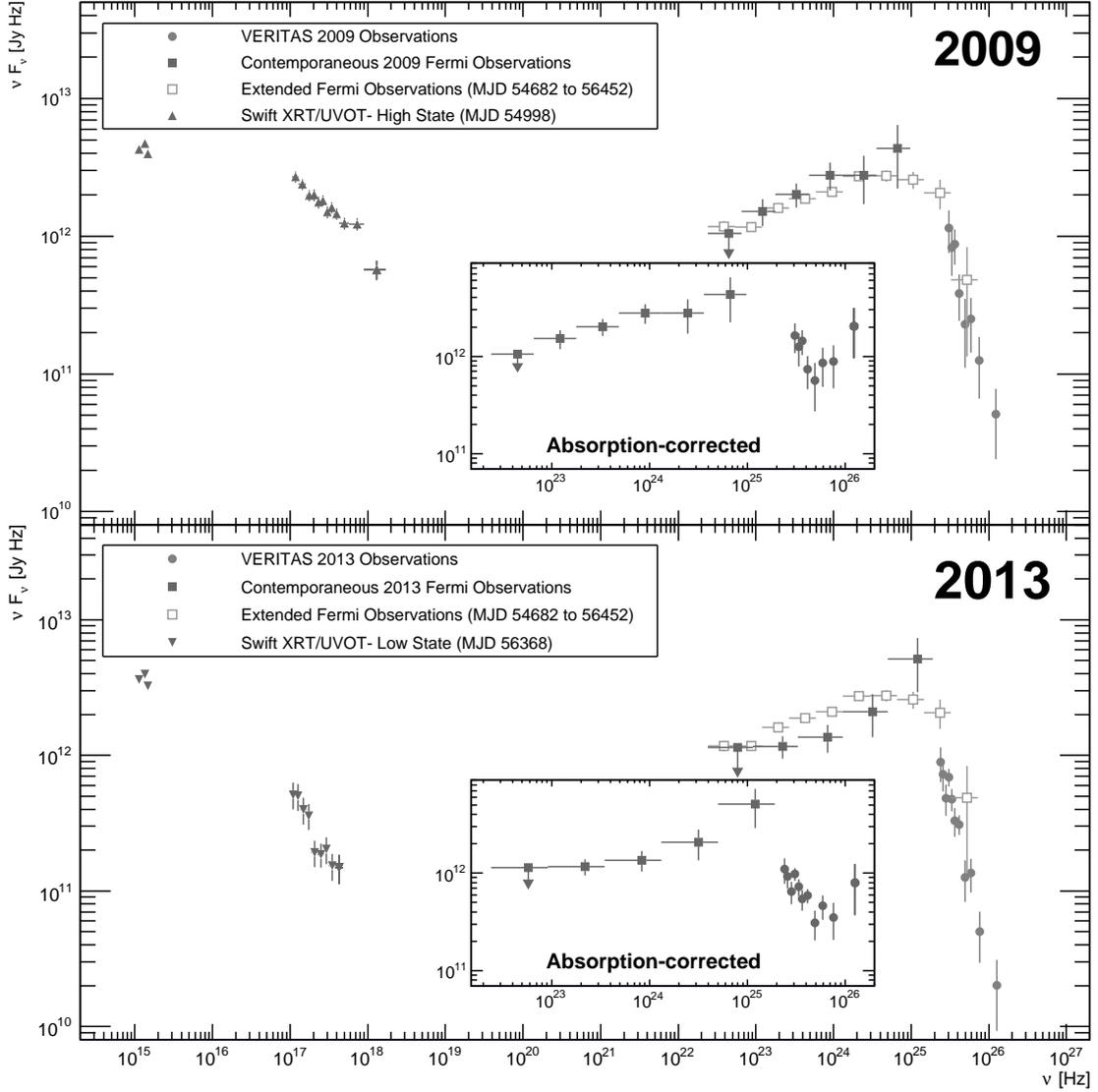}
\caption{Two broadband SEDs of PKS\,1424+240, corresponding to a relatively high (upper panel) and a low (lower panel) state. Within the inset, the VHE data are corrected for absorption using the \cite{gilmore2012} EBL model for $z=0.6035$.  The contemporaneous LAT data above 100 MeV are shown along with the spectral results from full LAT observations. The \textit{Swift} XRT and UVOT observations for relatively low and high states are also shown, after correction for absorption by the Milky Way column density. \label{fig4}}
\end{center}
\end{figure}

\rotate
\begin{scriptsize}
\begin{deluxetable}{cccccccccccccc}
\tabletypesize{\tiny}
\tablecolumns{14}
\tablewidth{0pc}
\scriptsize
\tablecaption{Summary of Gamma-ray Observations of PKS\,1424+240.}
\scriptsize
\tablehead{
\colhead{}    &  \multicolumn{9}{c}{VERITAS Results} &   \colhead{}   &
\multicolumn{3}{c}{\textit{Fermi} LAT Results} \\
\cline{2-10} \cline{12-14} \\
\colhead{Epoch} & \colhead{Exposure}  & \colhead{ON/OFF Region\tablenotemark{0}}  &\colhead{Excess}  & \colhead{Signal} & \colhead{Spectral Reconstruction} & \colhead{Index} & \colhead{Flux $>$120 GeV} & \colhead{Percent} & \colhead{$\chi^2$/DOF} &
\colhead{}       & \colhead{Index}   & \colhead{Curvature}& \colhead{Flux [0.1-300 GeV]}\\
\colhead{} & \colhead{[hr]}   & \colhead{Events}& \colhead{Events}  & \colhead{[$\sigma$]} & \colhead{Range [GeV]} & \colhead{$\Gamma$} & \colhead{[$\times 10^{-7}$ m$^{-2}$s$^{-1}$]} & \colhead{Crab [\%]} & \colhead{} &
\colhead{}        & \colhead{$\alpha$}   & \colhead{$\beta$ [$\times10^{-2}$]} & \colhead{[$\times 10^{-8}$ cm$^{-2}$s$^{-1}$]}}
\startdata
2009\tablenotemark{1} &28.5&3264/19635  & 423&8.5 &120-750 &3.8$\pm$0.3 &2.1$\pm$0.3 &4.6 &3.2/6& &1.73$\pm$0.07 &\nodata&8.3$\pm$1.3 \\
2011\tablenotemark{2} &14.6  &4189/24792 &540 &8.1 & 115-375& 4.3$\pm$0.6& 2.1$\pm$0.3&4.6 &7.3/6& &   1.79$\pm$0.08  &\nodata&7.8$\pm$1.2 \\
2013\tablenotemark{3} &66.8  & 12869/76307&1675 &14.4 &100-750 &4.5$\pm$0.2 &1.02$\pm$0.08 &2.2&7.5/9 & &   1.77$\pm$0.09 &\nodata&6.3$\pm$1.2 \\
Contemp.\tablenotemark{4} &109.9  &20322/120734&2638& 18.1&100-750 &4.2$\pm$0.3 &1.30$\pm$0.08 &2.8 &21.2/9&  &1.77$\pm$0.05  &\nodata& 7.7$\pm$0.7 \\
Full\tablenotemark{5} &  \nodata &\nodata& \nodata& \nodata & \nodata&\nodata &\nodata & \nodata &\nodata&& 1.64$\pm$0.06&(2.7$\pm$0.8) &7.37$\pm$0.04 \\
\enddata
\tablenotetext{0}{\scriptsize Gamma-ray signal calculated according to \cite{lima}, with ratio between ON and OFF region sizes of $\alpha$=0.167, 0.167 and 0.200 in 2009, 2011 and 2013, respectively.}
\tablenotetext{1}{\scriptsize MJD 54881-54888, 54937-54943, 54968-54982, 54994-55003.}
\tablenotetext{2}{\scriptsize MJD 55595-55604, 55620-55629, 55647-55662, 55677-55689, 55706-55711.}
\tablenotetext{3}{\scriptsize MJD 56334-56341, 56358-56374, 56384-56400, 56413-56428, 56441-56447.}
\tablenotetext{4}{\scriptsize Contemporaneous: includes all 2009, 2011 and 2013 epochs summarized above.}
\tablenotetext{5}{\scriptsize \textit{Fermi} LAT data between MJD 54682 to 56452.  Data fit with a log-parabolic model which includes absorption by the \cite{gilmore2012} EBL model. }
\end{deluxetable}
\end{scriptsize}

\end{document}